# *SugarChain*: Blockchain technology meets Agriculture - The case study and analysis of the Indian sugarcane farming

**Naresh Kshetri*[1], Chandra Sekhar Bhusal[2], Dilip Kumar[3], Devendra Chapagain[4]**

[1]Lindenwood University, USA, NKshetri@lindenwood.edu,
[2]Federation University, Australia, Chandra045Bhusal@gmail.com
[3]United University, India, DilipKumar.phdcs21@uniteduniversity.edu.in,
[4]Tribhuvan University, Nepal, DevCpgn@gmail.com

**Abstract.** Not only in our country and Asia, but the agriculture sector is also lagging all over the world while using new technologies and innovations. Farmers are not getting the accurate price and compensation of their products because of several reasons. The intermediate persons or say middlemen are controlling the prices and product delivery on their own. Due to lack of education, technological advancement, market knowledge, post-harvesting processes, and middleman involvement, farmers are always deprived of their actual pay and efforts. The use of blockchain technology can help such farmers to automate the process with high trust. We have presented our case study and analysis for the Indian sugarcane farming with data collected from farmers. The system implementation, testing, and result analysis has been shown based on the case study. The overall purpose of our research is to emphasize and motivate the agricultural products and benefit the farmers with the use of blockchain technology.

## I. INTRODUCTION

Sugarcane is a widely grown crop in India. Even sugarcane plants spread in most of the sub-continent in India. In India, sugarcane grows over 49.18 lakh hectares and India is set to become the fourth greatest growing financial system in the world (with USD 4.9 trillion in 2024 as Germany) next to the USA (25.3 trillion), China (20.6 trillion) and Japan (5.6 trillion) [1]. Sugarcane (cultivated in more than 109 countries) is the primary root of sugar except the sugar juice is used for producing white sugar, brown sugar and jaggery (also called Gur). Other than these, the main products (by-products) of the sugarcane production are bagasse (food products) and molasses (viscous liquid produced during sugar production from raw juice) [2].

Bagasse is primarily used as fuel. It is also used for invention of compressed fiber board paper, plastic and other. Molasses is castoff in distilleries for developing of ethyl alcohol, butyl alcohol, citric acid etc. More specifically, we can say that it delivers employment to millions of individuals directly or indirectly. But actually, there are various problems faced by farmers of India like varieties, manufacture technologies, farm outfits, climate variation, artifact development, value accumulation, marketing etc. Actually, the former government has taken a lot of decisions. Perhaps there is no improvement in earning for farmers in India and needs to shift from simple farming to more efficient, justifiable, and productive farming. India is an agricultural kingdom and agriculture is only ~16% of GDP in India but the prime sector for occupation [3].

In this paper we have explored the possibilities of using the blockchain technologies in sugarcane farming so that automation in sugarcane agriculture can improve production quality, quantity and efficiency. Farmers of sugarcane can create a global market and Blockchain technology can be one of the important pillars for this evolution of technology in agribusiness in addition to cybersecurity and cyber defense [4]. As blockchain technology is based on transparency, security and authenticity, it may be a great value chain to better understand possible emerging technologies in the field of agribusiness.

*Corresponding author, Dept. of Math, CS & IT, Lindenwood University, MO, USA.

Blockchain is the technology after the world's common cryptocurrency i.e., Bitcoin [5]. The main advantage of the blockchain technology was the probability of allowing information interchange between the parties without physical occurrence. In dissimilarity to the client-server architecture, blockchain technology is grounded on peer-to-peer (P2P) architecture [6]. The power or ability here doesn't relax on a sole computer rather the power rests on all the contestants of networks. Distributed peer-to-peer environment of blockchain-based technologies has no sole point of catastrophe. It uses the popular technical advancements like hashing, digital signatures, consensus mechanisms and difficulty correction. The consensus algorithm engaged by blockchain has proved the security, scalability and confidentiality of the podium via operating on 100% verification and 0% trust [7] [8].

The existing agriculture industry lacks the efficient food traceability method that can trail and monitor the entire life cycle of food making that includes the processes of food raw substantial cultivation/breeding, processing, transporting, warehousing and marketing that involve a bulky number of disloyal business parties [9]. Food traceability attempts to record, store and transfer sufficient data about food, feed, food-producing creatures or substances at all the phases of the agricultural supply chain. Such statistics is crucial to check the product for security and eminence control and can be sketched upward or downward at any time [10]. Lack of trust among the countless actors of the supply chain and distrust of customers about the quality of product are other major issues seen in the traditional agricultural supply chain. Blockchain as a circulated ledger where data is immutable and can be accessed by all the participating actors promises to solve the major problems of current agribusiness. Following are approximate of the profits of using blockchain technology in agriculture:

*1. Product Traceability:* Every information is recorded at each stage of the product's journey to the customer's table and is visible to all participants. The customers are able to trace the source and quality of the product. Such admission to the product's information ensures the consumer about the safety and quality of the products they consume [10]. The immutable and permanent stored information in the blockchain enables tracing much quicker, efficient, and easier [11].

*2. Building Customer Trust:* The immutable information stored in blockchain can be accessed by the customer using their mobile devices that enable them to trace the critical information like its origin, how they are grown and processed. This reliefs to enhance the customer's trust and sureness in the product they ingest [10].

*3. Enhance Supply chain:* Blockchain-enabled agriculture supply chain comforts to deliver real-time information to all supply chain participants improving the efficiency and transparency of the agricultural supply chain. The birthplace and quality of the product can be proved at any point thus decreasing and removing product waste in their process. The flow of immutable, real-time, and quality info advantage to accomplish the inventory and charge dynamically [10].

*4. Ensure Food Quality:* Blockchain along with IoT enables us to capture real-time information like product's condition (humidity, temperature) during the transition between various actors. Such information helps to ensure the quality of the product at every stage of the supply chain [11].

*5. Increase Efficiency in Supply Chain:* The product's information that can be accessed by all the participating actors not only build trust among each other but also help to make supply chain efficient and effective. The issue of timely payment to different stakeholders in the supply chain also shows a major role in well-organized supply chain. The smart contract as part of blockchain qualifies it to be a reliable means of distribution of payment among various actors in an efficient and well-timed manner as conditions of each contract are met. No actors have to wait for their payment as payments are initiated automatically from the fund that has been authorized by the bursars when certain conditions are met [10] [11].

Our study and analysis is arranged as follows. In *Section II*, we present a brief outline of background study, historical works and try to identify the necessity and tradition of BCT in the agriculture sector. We then discuss in *Section III*, about the existing methods and models used (including their impact and possibility for Indian sugarcane farming) followed by the case study (based on the primary data i.e., questionnaire developed for farmers) in *Section IV*. We have recommended the BCT model to help farmers for their actual pay and efforts and to automate the process with high trust in *Section V* with system implementations and testing of our research work in *Section VI*. In *Section VII*, we presented the analysis of our study and outcomes along with the conclusions and possible forthcoming scope in *Section VIII*.

## II. LITERATURE REVIEW AND RELATED WORK

In [10], **N. Kshetri et. al.** (2021), surveyed many former works in the field of BCT and its applications linked to agribusiness. The paper also deliberates the advantages and situations through current improvement of the agribusiness zone. It is witnessed that the transparency and trust of BCT with its widespread network architecture (using blocks of data and information) can profit agribusiness in several magnitudes. The authors have also projected a blockchain-based agribusiness model (that is of unique nature) for info transparency and refining the efficiency of the communal sector in absence of third parties in the network. It looks that the ground-breaking technology, BCT, can help several sectors including the agribusiness zone too. Customers and farmers (along with all agribusiness consumers) both are profited with the use of BCT-based agribusiness. The effectiveness in the supply chain can be enhanced and delivered real-time to all affiliates that can alter the inventory and charge of the product.

In [12], **F. Miatton and L. Amado** (2020), introduced the notion of Commodity Fairness Index used to size the inequality, or economic imbalance in a commodity value chain, and estimates it in the case of Colombian coffee. Coffee is one of the most broadly consumed beverages in the ecosphere and internationally merchandised commodities, but the coffee value chain is impervious and disturbed. The application of BCT to the coffee manufacturing enables inclusive business models that rewards quality and tough work, and in turn converts into greater trust, confidence and fairness across the complete industry as well as among end buyers. The authors also described the system architecture of a network application built upon Hyperledger Fabric that surfaces the way to improve coffee farmers' survives by bringing transparency and traceability into the whole value chain and refining its fairness as an effect.

In [13], **H. Fang et. al.** (2020), provided a review to study both techniques and applications of BCT in the agricultural zone. First, the practical elements, with data structure, cryptographic methods, and consensus mechanisms are clarified in detail. Second, the prevailing agricultural blockchain applications are classified and reviewed to reveal the use of the blockchain techniques. Increasingly, BCT is drawing significant attention in numerous agricultural applications. These applications could please the diverse needs in the ecosystem of agricultural products, perfection of contract exchanges and transaction efficiency. The authors also provided widespread platforms and smart contracts to display how practitioners use them to develop agricultural applications. Thirdly, the authors also identified key defies in agricultural systems, and debated the efforts and solutions to challenge these problems with an upgraded food supply chain (in post COVID-19 epidemic) as an sketch to demonstrate an effective practice of BCT.

In [14], **M. D. Borah et. al.** (2019), implemented an easy web-based stage in Agricultural Supply Chain Management (SCM) using BCT, which is a decentralized safeguarded system to get transparency, enhanced product value. Blockchain plays a vivacious role in FARMAR to track and trace the source of food products in the food supply chain. SCM is a crucial business process in all domains of the economy. SCM uses precise processes to connect from producer to consumer necessity through a chain. Additional advantage of using BCT in FARMAR is security where hacking or altering of the existing data is difficult by any intermediary. As an add-on to this process, IoT devices (Mobile phone-based Android apps) are used to modernize the real time quality and transfer time of the product in FARMAR. It is integrated for upgraded traceability and usability of the products in the supply chain. The writers also believe that the FARMER aims to succeed these goals by developing a web application where FARMAR makes a value chain of integrity from farmstead to fork by using BCT.

In [15], **Z. Hao et al.** (2019), proposed an automatic commercial transaction mechanism i.e., consortium blockchain model where food can be exported independent of a favorite third party. The novel Food Trading System with COnsortium blockchaiN (FTSCON) improves trust and security concerns in transactions. FTSCON uses consortium BCT to set authorization and authentication for altered roles in food transactions, which meet the defy of the privacy protection of multi-stakeholders. Orthodox food trading stages face several issues, such as hurriedly to find trading objects and shelter the reliability of transaction data. With e-commerce developing quickly, food trading has also recently moved to the online domain. Blockchain has reformed many industries owing to its robustness, decentralization, and end-to-end credibility. Security analysis displays that FTSCON progresses transaction security and privacy protection by announcing a smart-contract life-cycle supervision method. Experiment outcomes based on a series of data signpost that the proposed algorithm can triumph profit improvement of traders.

In [16], **V. Sudha et al. (2021),** recommended a blockchain based supply chain management system that would crack all the problems of outdated and poor supply chain administration in India. The paper highlights the major glitches faced by Indian farmers: (i) absence of facility to stock their products, (ii) incapable to monitor the product's standing and sell them with income. The proposed system is supposed to provide transparency about the product's status while preserving a good connection among producer and consumer. Authors present the architecture of the proposed system where product information is composed from the farmers using the application beforehand they are conveyed from the farmer's place and this information is confirmed by the smart contract. Once all the information is established correct, it is warehoused in the blockchain network. Different sensors (IoT agents) are located in the storage place and transportation vehicle to measure temperature, humidity and existence of chemicals. All the complete information recorded by sensors are stored in a blockchain that lets together farmers and consumers know the status of the products at many stages of the supply chain. The system also aids to track the varying market price and integrate the recent market price in the blockchain.

In [17], **Q. Lu & X. Hu** (2018), developed a model called OriginChain from the case study for Product Traceability. OriginChain restructures the service provider's present traceability system by interchanging the central database with a blockchain. OriginChain offers transparent tamper-proof traceability data, enhances the data's availability, and automates regulatory-compliance checking. The authors have implemented and tested OriginChain under realistic circumstances employing the user's traceability information. OriginChain currently employs a geologically distributed private blockchain at the traceability service provider, which has branch offices in three countries. The greatest visions that authors learned is the design *(impact on cost, if it is public blockchain because of more lines of code)* and adaptability *(quality attribute required by many industrial projects that are dynamic, also means the smart contract to be modernized by a number of authorities beyond the threshold demarcated in the factory contract)* of blockchain-based systems.

In [18], **T. Rocha et al.** (2021), proposed SmartAgriChain that anticipates to implement a supply chain and certification system based on Hyperledger Sawtooth that will be proficient of identity management, hierarchical users/organizations, significant scalability, little costs, little energy consumption and compatibility with legacy schemes. The proposed result does not compromise currently existing features, but it will, however, permit all the actors to take part in the agri-food supply chain system and persistently monitor its actions. Management of certification issuance and product counterfeit proofs in the agri-food supply chain are very serious and reaching glitches nowadays. The currently existing management systems for this process are either obsolete or have significant issues when it comes to security, trust, traceability, management or product certification. The introduction of BCT, due to its core properties, has the potential to crack identity, ownership, data temper, traceability, and certification issues. The writers also explored and explained the system design and architecture in detail as well as a price projection based on the total of nodes of the distributed system.

In [19], **Gunasekera, D., & Valenzuela, E. (2020)** analyzed the concept of economic effect due to blockchain adoption in Australian grain. Authors has further refined the concept of productivity gain quantified using a Global Trade Analysis Project (GTAP) model, which is a broadly used computable general equilibrium tool for analyzing the international economy and undertaking the illustrative scenarios of (i) Blockchain use (grains) scenario: (ii) Blockchain use (finance) scenario: and (iii) combined scenario: combining scenario 1 and scenario 2. From the ancient trends in productivity improvements in key industry sectors, adoption of blockchain technology in Australian grain sectors will increase productivity by five per cent to ten per cent over ten years (2020-2030). The analysis shows increase in productivity due to reduction of transaction cost after adoption of blockchain technology as a 'distributed ledger technology' that empowers the quick settlement of payments to grain producers as a possible productivity gain in business transaction services sectors that provide services to the agriculture sectors.

In [20], **Lin, J., Shen, Z., Zhang, A., & Chai, Y. (2018)** recommends a blockchain technology and IoT based food tractability system based which is confidential and self-organized. Writers described the architecture of the proposed system that comprises the traditional ERP legacy system and a new IoT system connecting all parties in agri-business. The system is a computer-generated blockchain network consisting of dual types of nodes where one is furnished with all functionalities of blockchain node and additional is the thin node which is just a simple payment verification (SPV) node that only authenticates the compensation and stores transactional statistics. IoT technologies eradicate human intervention by exchanging manual recording and verification as possible. All actors including consumers would be able to entrance the data stored in the system and authenticate them using their smart phone thus growing trust among the actors. Further writers plan to implement smart contracts that would relief law executors for problem ID and well-timed processing.

In [21], **Xiong, H., Dalhaus, T., Wang, P., & Huang, J. (2020)** examined the four uses of blockchain technology in agriculture and food sectors: food supply chain, agriculture insurance, smart farming and transaction of agricultural products. Decentralized crop insurance grounded on blockchain technology and smart contracts enable automated payouts to compensate the farmer can be revolutionary in agriculture insurance. Smart farming can be achieved with a combination of IoT and blockchain technology that serves to store data and information generated by all the actors in a distributed database providing access to all participants ensuring trust and transparency. Blockchain with characteristics of decentralization, security and transparency help to address the existing difficulties in food supply chain for instance food traceability, food security and trust, supply chain inefficiency by making it probable to track all the product information of food quality and safety in the complete supply chain. The author highlights the challenges of e-commerce trade of agricultural products such as information insecurity, cash on delivery and high operating costs which can be solved by practice of blockchain technology. Blockchain (i) provides information security by encryption mechanism that provides the authentication requirements, (ii) enhance supply chain management by lowering the information sharing cost among the actors, (iii) provide digital payment solution with cryptocurrency and (iv) build customer confidence by letting customer access all the product information in transparent manner.

In [22], **Kamilaris, A., Fonts, A., & Prenafeta-Boldύ, F. X. (2019)** surveyed the adoption of the blockchain technology and its impression in agriculture and food supply chain. Authors have analyzed the challenges and potential of some of the ongoing projects and initiatives. Food security can be achieved with blockchain technology that provides transparent details, creating records and resources verifiable and accessible to respond more promptly and efficiently in case of emergencies. One example is *Blockchain for Zero Hunger (2017),* where digital food coupons was distributed via Ethereum-based blockchain to Palestinian immigrants in the Jordan's Azra camps. Blockchain can address the issue of food safety via traceability of products at each stage in the supply chain that enables to guarantee good hygiene conditions, identify frauds, risks and unhygienic products in the early stage. Some of the initiatives for food safety are *Walmart* and *Kroger* that adopted blockchain technology for the supply chain Chinese pork and Mexican mangoes. Another impact of blockchain technology is the food reliability which is about swapping the food reliably where blockchain enables each actor of the supply chain to exchange all details of origin of the product. *Downstream* Beer Company used the blockchain technology where every detail of the beer is recorded in the blockchain which can be accessed by customers using their smartphones. Similar companies who adopted blockchain technology for food integrity are ecommerce platform *JD.com*, *Grass Roots Farmers' Cooperative*. Another impact of blockchain technology includes support to small farmers by forming cooperatives of farmers that enable farmers to obtain a larger portion of the value of the crops they grow by joining cooperatives. Blockchain technology is incorporated in various waste management and environmental awareness, for instance *Plastic Bank* was established to reduce the plastic waste by rewarding the blockchain-secured digital token for those who bring plastic bags to bank recycling centers. The article concludes with blockchain technology having huge potential in the future once issues like accessibility, technical and design issues, policy regulating the blockchain technologies are resolved.

In [23], **Demestichas, K., Peppes, N., Alexakis, T., & Adamopoulou, E. (2020),** presents a summary of how blockchain technology implementation has enabled the trace of agri-food products. Author also presents a short-term explanation about blockchain's architecture, smart contracts, consensus methodology and kinds of blockchains. Author's chatted different existing blockchain based frameworks in blend with IoT and smart contracts whose central aim is food safety and food security. Blockchain delivers an immutable distributed ledger with an encryption mechanism to segment every product's information at every stage of the supply chain with each patron. Authors also highlight some of the businesses who have built blockchain based systems for tracking and tracing the agri-food products. Some of them which are IBM Food Trust to trace the origin of the Chinese pork products and mangoes, Provenance for tracing fish products, AgriOpenData for tracing whole agri-food, AgriDigital for refining whole grain supply chain by making it easy, humble and secure to connect farmer and purchaser. Authors also present some of the defies for blockchain implementation despite the technology gaining more and more space in the supply chain.

In [24], **Torky, M., & Hassanein, A. E. (2020)**, examines the impact of integrating blockchain technology and IoT to develop smart applications to be used in agriculture and also proposed a model that could address some of the major defies in IoT based agricultural systems. The author also discusses the challenges of developing the blockchain-IoT system for agriculture. The author reveals that integration of blockchain and IoT would be a great contribution to revolutionize the digital transformation in various domains including agriculture and highlights top five blockchain and IoT predictions by 2030 .In the proposed model, IoT peers cooperate and shape trust over the blockchain network. The IoT transactions are broadcasted on the network based on the blockchain protocols. The authors proposed the hybrid design form that integrates IoT, Cloud computing, Fog computing and Blockchain where blockchain can be used to function as data warehouse and transaction monitoring and verifier for dissimilar heterogeneous fog networks which are controlled by cloud. Blockchain can provide solutions for (i) sensing problem of IoT devices (ii) energy consumption in IoT devices, (iii) network complexity problem of IoT devices, (iv) bandwidth and latency problem of IoT devices communications and (v) limited data storage problem of IoT devices. Authors also discussed some blockchain's technical challenges of storage and scalability, forking, latency and throughput.

In [25], **Revathy, S., & Priya, S. S. (2020, September)** proposes a Blockchain-based Producer-Consumer Model (BPCM) which permits the farmer to trade their commodities directly to the buyer while avoiding the inter agents using the smart contracts that permit the farmer to add more profit. The writer also examines the blockchain features along with its application and discusses the profit of farmer's direct marketing. In the outdated producer consumer model, the farmer or producer has to sell their agri-product to the retailer and the retailer sells the same to the consumer by growing the price. Here inter agents get more profit as farmers have no control in setting the amount. The proposed BPCM is developed using Ethereum, a communal blockchain where farmers and consumers are provided with an exceptional identity, and all are connected to the blockchain network. The nodes willing to link the BPCM have to verify its identity using Proof of Work (PoW) consensus where nodes will not be added to the network if the node fails to prove its individuality. The model limits the transactions between the consumers' nodes that will not let any inter agent who intend to pretend the consumers using the smart contract. Here the consumers can purchase the products directly from the farmer at a practical price. The transaction can be originated by either producer (farmer) or the consumer with their individual identity that generate the block for the transaction. The newly formed block is then broadcasted in the network to all the nodes for authentication. The smart contract is used for the authentication of the transaction where it checks whether the transaction is requested amongst the farmer nodes and consumer nodes or between the consumer nodes. If the transaction is amongst farmer node and consumer nodes, smart contract will execute the transaction as lawful. The transaction request from consumer node to consumer node is blocked allowing the farmer to sell their products to 'n' consumers while limiting the middleman in the identity of the customer to earn the revenue. The Proof of Work (PoW) consensus algorithm is used to guarantee that all the nodes in the BPCM approve to the transaction. Finally, when the transaction is confirmed, the block is added to the chain where the transaction gets implemented. The blockchain ledger is restructured with new transactions and each node maintains a replica of the transaction. In this model the transaction is not confirmed if (1) the node failed to verify its identity and (2) the transaction initiated is amongst the customer nodes. The proposed model aids a unique model where farmers directly can sell their product to buyers where they have control on the price and consumers also get product at practical price while maintaining direct relationship with the farmer.

In, [26] Papa, Semou, (2017), presents how exchanges are recognized between farmers and cooperative, between cooperative and the transformer or amongst the transformer and the dealer. As they are not present at the time of interchange none of the actors has admission to all of the transaction. So, this transaction can be decentralized using block chain technology operating without a vital body. This technology concentrates on creating direct relationships while growing confidence and visibility into the movement of goods. Papa Semou also termed certification and traceability in agriculture. These traceability techniques will aid to capture, store and manage all the data of the product. Finally, the blockchain assurances to not only be limited to the simple recording of transactions but also implementing computer programs.

## III. EXISTING METHODS & BUSINESS MODELS

The first step of using BCT is always to improve the equilibrium and fairness of the respective industry to convey transparency. The system design and communication pattern in such models / architecture is very important in order to enable transparency and traceability into the total value chain. A web-built app upon the Hyperledger Fabric blockchain framework is described where message movement starts with a user interacting with a web-based app whichever by desktop or smartphone, introducing the idea of Commodity Fairness Index (CFI) [12]. The request to fetch / write the data, if allowed, is sent to the REST API responsible for generating a link amongst server and client. API (that uses queue manager to store request) acknowledges request and assigns a matchless key that will be used to receive the blockchain's wish later on. The examples of requests can be to register a farmer, query history from blockchain, update the status and location of a shipment as it moves along the supply chain. The message flow passes through API, Nodes, Certifying Authority, Ordering Service, Distributed memory-key database, and at the end web/mobile app accepts the response to its new request via REST API after querying the distributed ledger expending the unique key.

The case of the agricultural supply chain shows the farmers are underprivileged of their fair part and the intermediaries. To eradicate the middleman in supply chain management, BCT can be helpful by the proposed method - FARMAR [14]. This process involves of a series of steps where each and every IoT device is kept for tracing and all data is added to the blockchain network via a shared ledger. The experimental setup and tools used for the FARMAR are BigchainDB, Linux (UBUNTU), Python 3.6, and Monit. BigchainDB (version 2.0b9) is a database with blockchain features that has great throughput, low latency, powerful query functionality, decentralized control, immutable data room, and built-in asset support. Linux (Ubuntu version 16.04 and above) is the OS for easy, fast installation, and setup with good community support. During the development and testing phase, writers have used a fresh version of PYTHON and PIP. The writers have used Monit's (Watchdog) for system observing and fixing errors because one cannot make sure that all the servers are running brilliant all the time.

Trading of food also shifted to the online domain with e-commerce growing day by day. BCT has reformed many industries due to its robustness, decentralization and end-to-end credibility. To improve trust in security issues and transactions, a novel Food Trading System with Consortium Blockchain (FTSCON) was proposed [15]. It practices consortium BCT to set permission and authentication for altered roles in food transactions, which meet the test of the privacy protection of multi-stakeholders. The architecture of FTSCON includes two entities: User node and Scheduling node. The consortium blockchain is made up of three fragments: Block containing the transaction data, Consensus mechanism and Smart contract. The consortium blockchain is a P2P model and the rights and compulsions of all peers in the P2P network are the equivalent. The algorithm of optimized transaction combination is intended for the purpose of serving users find suitable transaction objects. It can choose the optimized swapping portfolio for buyers. The virtual double auction mechanism is used to eradicate competition. Moreover, a smart-contract life-cycle management method is presented, and security analysis shows that FTSCON progresses transaction security and privacy protection. Experimental results based on a sequence of data indicate that the proposed algorithm can triumph profit improvement of merchants.

There is a significant advantage of employing BCT with IoT in the agriculture domain. A proposed system (blockchain based solution) is developed to keep way of the goods and manage the workflow [16]. Front end of the system is developed in JavaScript. Whenever farming goods are conveyed from a farmer's place, all the details about goods (such as size, color, nature, organic or inorganic, cultivation time, humidity, market rate etc.) are measured and kept in the blockchain. The model has a User Interface that connects to the Smart contract which leads to the creation of blocks in order to store in the blockchain. In the Blockchain, the transaction can be effectively created only if constraints specified in the applications are pleased. The transactions are the events produced by the nodes/parties in the blockchain. Before adding a transaction to a block, the transactions are confirmed by all the parties. A block is linked to the further block in the network by including the hash value. The commencement of a transaction is done by invoking the program in a smart contract. All the promises to be executed are in the form of condition statements. When all the requirements are fulfilled, the action followed by it takes place. When a transaction is effective, next it must be added to a block. The records preserved in the blockchain are immutable, values once stored cannot be changed by anybody, whereas any person in the SCM can see the documents.

Product suppliers and retailers usually want independent traceability service suppliers who are government-certified to examine the products throughout the supply chain. Suppliers need to receive certificates to display their products' source and quality to consumers and to conform to regulations. Retailers need verification of the products' origin and quality. Blockchains nurture continually because the data and code on them are immutable. The proposed model, OriginChain, now employs a geographically scattered private blockchain at the traceability service provider company, plot is to establish a consortium blockchain, which will include other organizations [17]. OriginChain has division offices in three countries and the plan is to establish a trustworthy traceability platform that covers other organizations, including government-certified labs, big suppliers and retailers that have long term associations with the company. Compared to a public blockchain, such a consortium blockchain (i.e., OriginChain a private blockchain) can accomplish better and cost less. Product suppliers or retailers accomplish product or enterprise information through the product-and-enterprise-management module. They access the information on the blockchain through a webserver introduced by OriginChain. After the traceability service provider authenticates an application from a product supplier, both parties sign a legal agreement about what traceability services are concealed. OriginChain creates a "smart contract" that represents the legal settlement.

| SN | Methods / Models | Author (s) | Title | Journal / Conference (Year) | Ref. |
|---|---|---|---|---|---|
| 1. | OriginChain (for product traceability) | Q. Lu & X. Xu | Adaptable Blockchain-based Systems: A Case Study for Product Traceability | IEEE Software (Volume: 34, Issue: 06, 2017) | [17] |
| 2. | Food Trading System with COnsortium blockchaiN (FTSCON) | Z. Hao et. al. | Novel Automatic Food Trading System Using Consortium Blockchain | Arabian Journal for Science and Engineering (2019) | [15] |
| 3. | Commodity Fairness Index (CFI) for Coffee Value Chain | F. Mittatton & L. Amado | Fairness, Transparency and Traceability in the Coffee Value Chain through Blockchain innovation | IEEE Conference on Technology and Entrepreneurship - Virtual (ICTE-V, 2020) | [12] |
| 4. | FARMer And Rely (FARMAR) | M.D. Borah et.al. | Supply Chain Management in Agriculture using Blockchain and IoT | Springer Nature Singapore (Book Chapter, 2020) | [14] |
| 5. | BCT-based model for SCM | V. Sudha et. al. | Blockchain-based Solution to improve Supply Chain Management in Indian Agriculture | IEEE International Conference on AI and SS (ICAIS 2021) | [16] |

Table-1: Summary of proposed existing methods / business models from 2017 - 2021 [12] [14] [15] [16] [17]

## IV. CASE STUDY FOR SUGARCHAIN

Our case study for sugarcane farming is based upon 15 questionnaires for the sugarcane farmers. The questionnaire is listed in the Appendix section of this paper. Forty sugarcane farmers from several states of India have participated in the case study and data collection process. We have used Google Forms to collect the primary data from the farmers. The results have been represented in various pie-charts and bar graphs making it easy for analyzing. The data collection and case study took almost 05 - 06 months due to various obstacles and hindrances (ongoing Covid and its variants in India, lockdown imposed by governments, lack of communication and language representation for farmers, nature of primary data etc.). Despite these, we were able to collect primary data from 40 farmers which represents the real and exact problems faced by sugarcane farmers.

Out of 40 sugarcane farmer responses, our first question, *"Q1: For how many years have you been doing sugarcane farming?"* has a mixed response of answers (67.5% stated more than 10 years, whereas 17.5% and 15% stated as 06-10 years and 01-05 years respectively). In our second question *(Q2: Is sugarcane farming the major farm for you?)* 80% have sugarcane as their major farm whereas 20% have indicated rice, wheat, grains, banana and other crops as their major farm beside sugarcanes. The third question of our questionnaire, *Q3: How long (in months) does it take approximately to get the sugarcane harvest?* has several answers but all answers are between a range of 8 months - 11 months. 75% of the responses said "Yes" for the fourth question, *Q4: Does your production cost is recovered by selling on a price set by the government?* whereas 20% and 5% responded as "No" and "Don't Know" respectively. Like the answer of Q3, the answer of Q5 *(At what rate do the sugar mills generally purchase your sugarcane?)* is also in the mixed response range of Rs. 3 - Rs. 12 (INR).

One major headache for sugarcane farmers can be clearly seen in the response to the sixth question *(Q6: Is the sugarcane affected by worms/viruses?)* 90% stated that their sugarcane is affected by worms/viruses, whereas 5% sugarcane farmers said "No", and the other 5% farmers said, "Don't Know". Another major issue for sugarcane farmers is the payment for their sugarcane purchased. Majority of the farmers in our survey (92.5%) stated that they only get paid after a few times (say 15 days, 1 month or more than that), whereas 7.5 % of the farmers only said they get their payment instantly (within a week or less than within a week). This problem of payment delay was asked to sugarcane farmers as our seventh question i.e., for Q7*: Did you get paid for your sugarcane sold instantly or after a few times?*

Beside the problems of sugarcane worms/viruses and late payment, farmers have pointed out various other problems in response to our eighth question *(Q8: Can you say one major problem, while cultivating and/or harvesting the sugarcane?)* of the questionnaire. We requested the participating farmers to point out one major problem at the time of cultivating or harvesting the sugarcane, in which 77.5% of the farmers have specified the problem. The common problems as raised by the sugarcane farmers are labor shortage (at both cultivating and harvesting times), water or irrigation problem, lack of manpower, expensive labor cost during loading and unloading of sugarcane (after harvesting), lack of seeds and fertilizers, lack of harvesting machines, lack of medicines for sugarcane, need to be moisture etc. Only 22.5% of farmers have denied saying or mention major problems for sugarcane farmers.

The above stated problems by almost 78% farmers (in Q8) isn't the end of the problem for farmers. Two major problems are clearly identified in response to Q9 and Q10. Finding the buyer (at their own effort and contact) for their sugarcane after harvesting it, is another dark side of sugarcane farming. In response to Q9 *(Do you need to find a buyer, or can you easily sell the sugarcane?),* 70% of the farmers agreed that they should find a buyer whereas 30% said we can easily sell the sugarcane without finding the buyer. Another serious issue outlined is at Q10 by 80% of farmers stating that weather/climate is affecting the sugarcane farming. Only 15% of the farmers replied with a "No" and 5% of the farmers said "Don't know" for the answer to the question *(Q10: Is weather impacting/affecting the sugarcane farming?).*

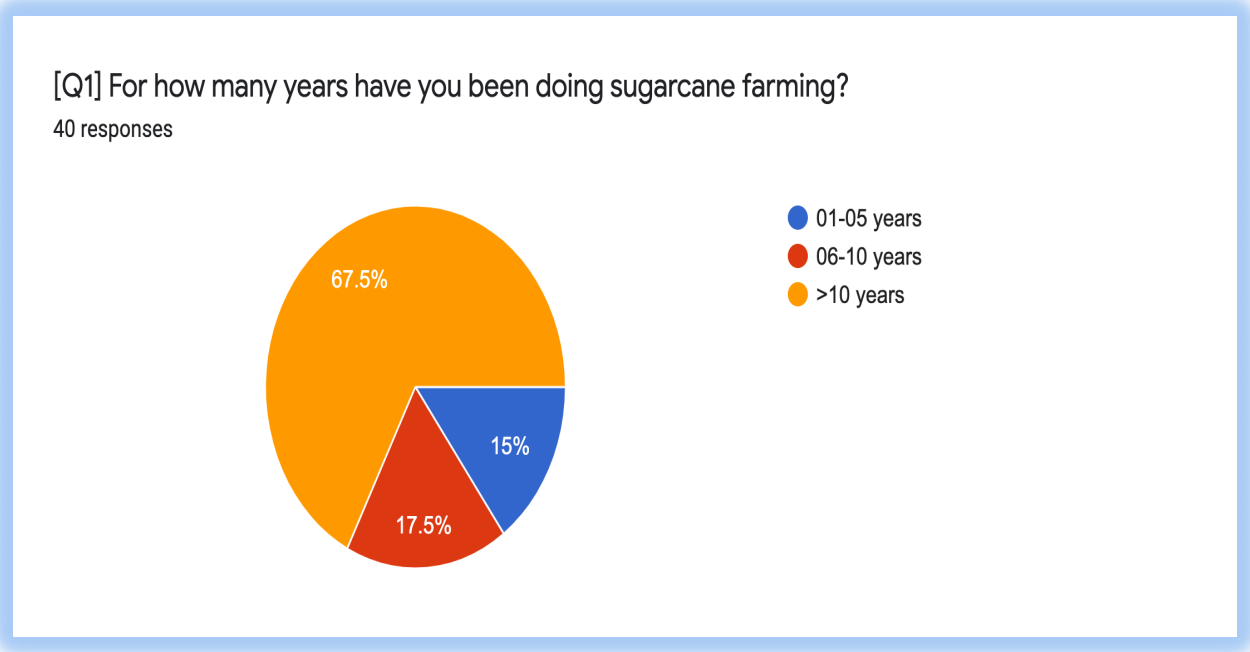

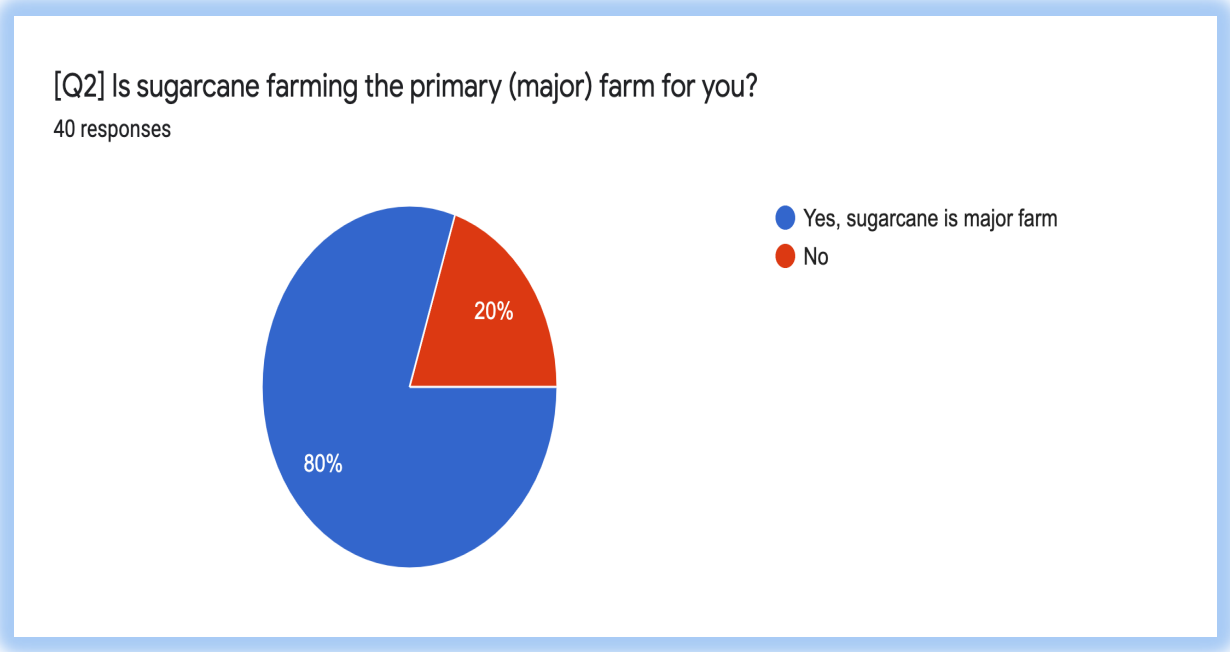

[Q4] Does your production cost is recovered by selling on a price set by the government?
40 responses

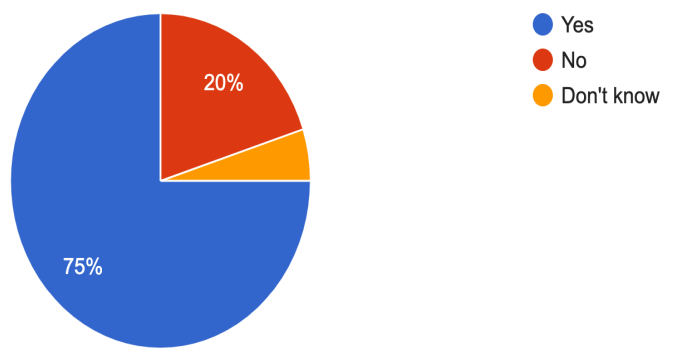

[Q6] Is the sugarcane affected by worms/viruses?
40 responses

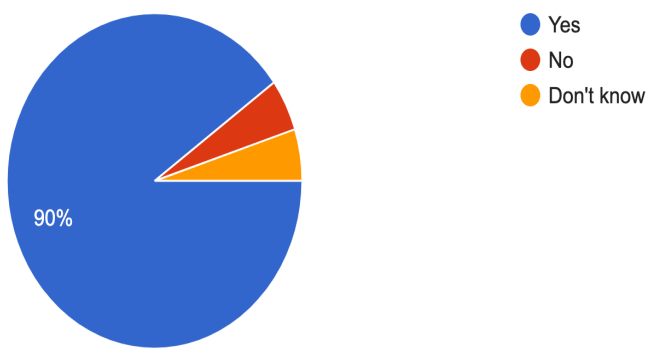

[Q7] Did you get paid for your sugarcane sold instantly or after a few times?
40 responses

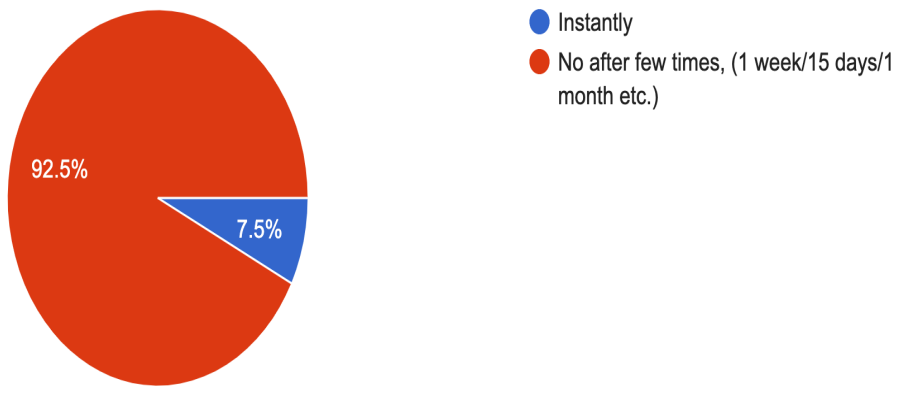

[Q8] Can you say one major problem, while cultivating and/or while harvesting the sugarcane?
40 responses

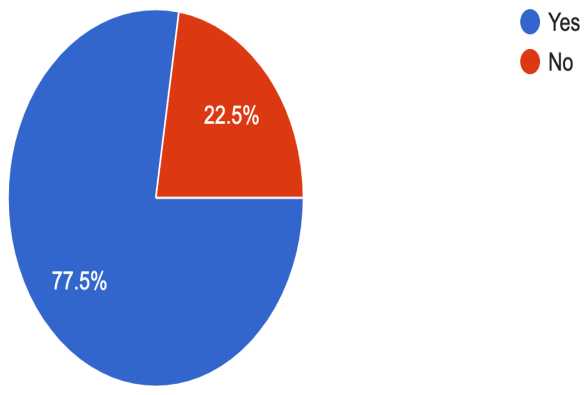

[Q9] Do you need to find a buyer, or can you easily sell the sugarcane?
40 responses

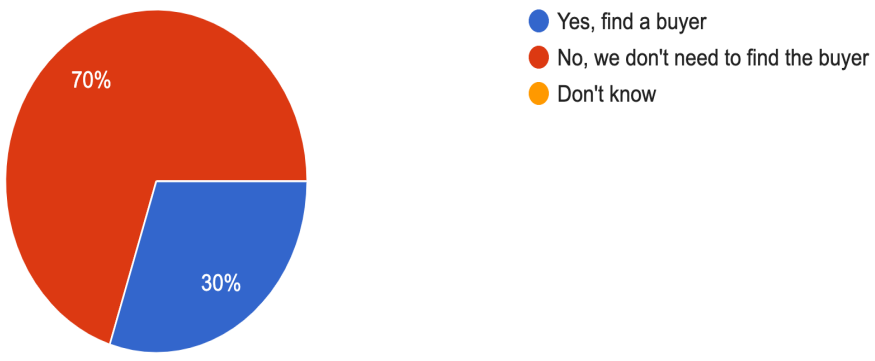

[Q10] Is weather impacting/affecting the sugarcane farming?
40 responses

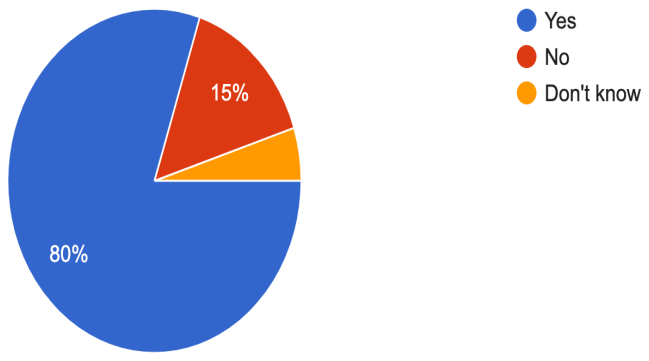

[Q11] Is the fertilizers and seeds for sugarcane available easily?
40 responses

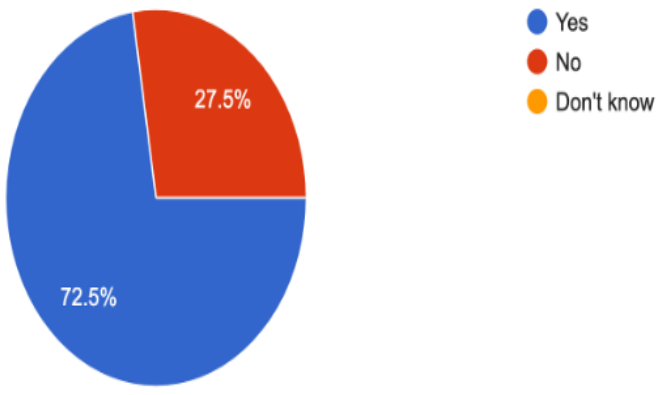

[Q12] How sugarcane reaches the factory for further processing?
40 responses

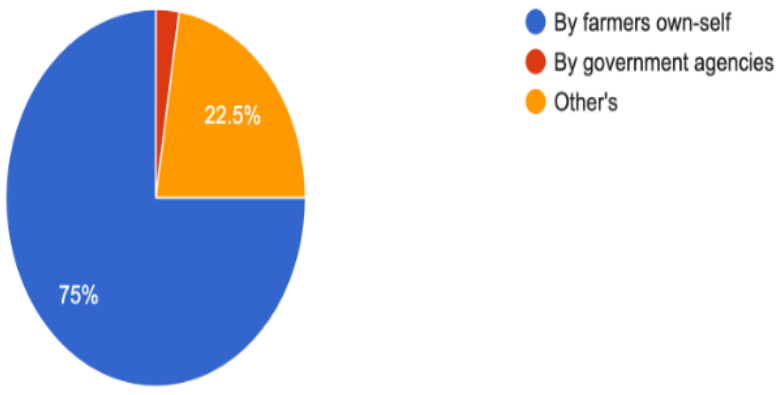

[Q13] Is there any impact of wild animals in sugarcane farming?
40 responses

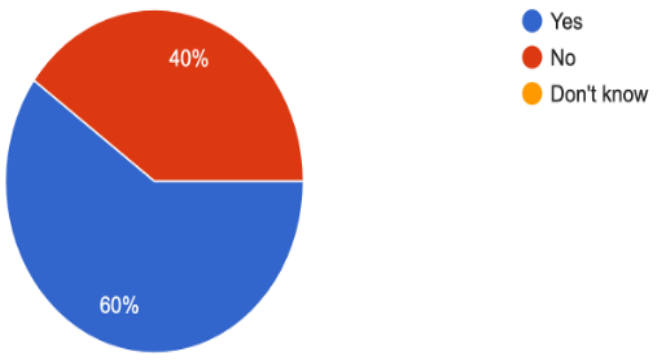

[Q14] Are there any problems faced by you (like transportations, gov. taxes or any other)?
40 responses

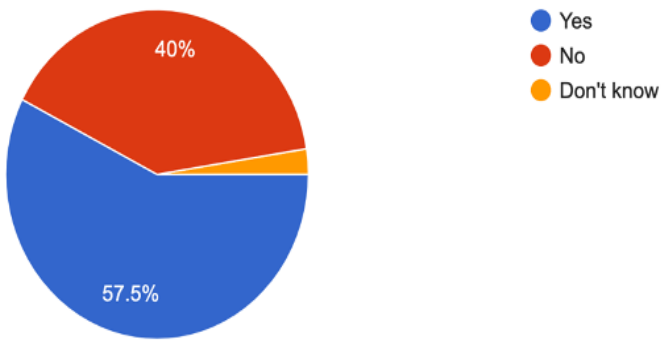

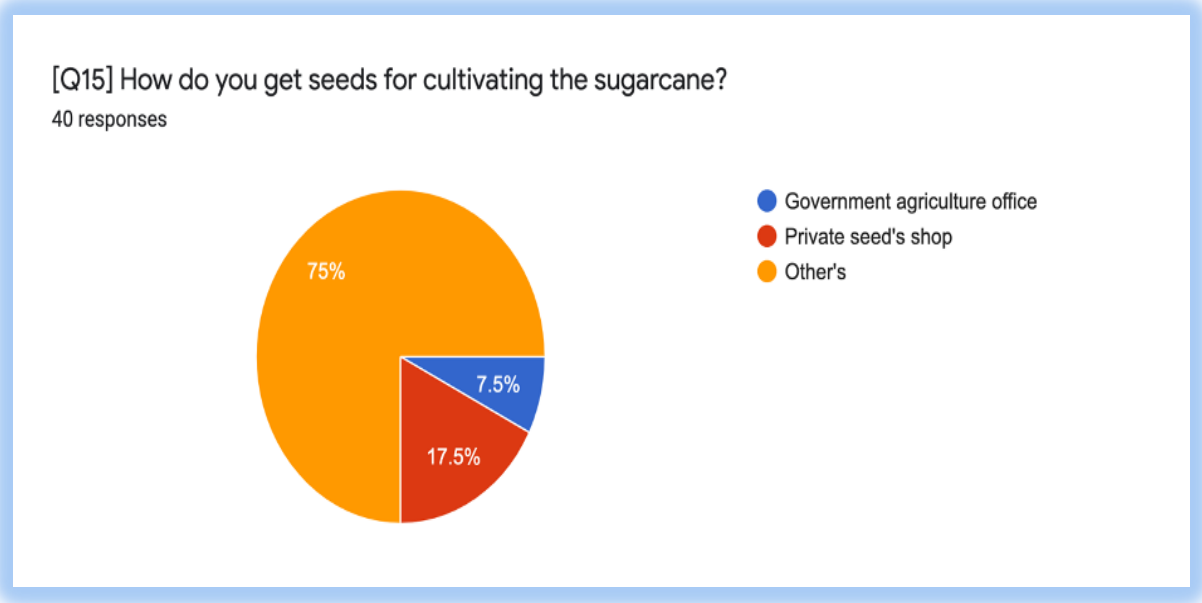

The availability of fertilizers and seeds for sugarcane is clearly mentioned by the farmers in response to the answer to the 11th question *(Q11: Is the fertilizers and seeds for sugarcane available easily?)*. Out of the three options provided in the Q11, 72.5% said they get the fertilizers and seeds available easily whereas 27.5% on the contrary said the fertilizers and seeds are not available easily. Although the fertilizer is available easily, the transportation of sugarcane is not available easily, which is mentioned in Q12 (*How sugarcane reaches the factory for further processing?*) of the questionnaire. 75% of farmers mentioned that they have to arrange the vehicles on their own from fields to the factory, but 2.5% only agreed that government help for such transportation and 22.5% stated "others" to reach the sugarcane to the factory.

The impact of wild animals (say, some wild cats, wild pigs, foxes, even leopards and tiger etc. apart from rats, snakes, and mice) also have an impact in the sugarcane farming as mentioned by farmers to the answer to 13th question (*Q13: Is there any impact of wild animals in sugarcane farming?*). Sugarcane fields are the homes to a number of insects and reptiles as mentioned by farmers, as many animals eat and damage the sugarcane crops. Almost 60% of the sugarcane farmers agreed that there is the impact of sugarcane farming whereas only 40% of the sugarcane farmers stated that there is no such impact.

The answer to the 14th question (*Q14: Are there any problems faced by you?*) about any other problems faced by farmers has pointed out several short-term and long-term difficulties of sugarcane farmers. 57.5 % of farmers have mentioned problems faced whereas 40% mentioned "no" problem, but 2.5% of farmers said "don't know" about that. Despite the lack of fertilizers, technical experts, late payments of sugarcane, farmers have pointed out transportation (truck/tractor) and labor shortage problems. Farmers also mentioned the issue of stuck vehicles (tractor / truck) after loading the sugarcane due to unpaved and narrow roads in many villages and districts. The last question of the case study is the 15th question (*Q15: How do you get seeds for cultivating the sugarcane?*), which says that only 7.5% of farmers get seeds from government offices. Majority of the farmers depend on either private seed's shop (17.5%) or other ways (75%) to get the seeds for sugarcane.

## V. PROPOSED ALGORITHM & FLOWCHART

The Algorithm A1 (SugarChain) describes a high-level process of user registration and login to the SugarChain. If the user is already registered, then he/she can login to the system. Upon the successful login, and user session is created. The algorithm also shows the process for password recovery.  If the user is new then he/she has to provide details (name, id, email, and phone) as input to register. Once a user is registered successfully, a userID (public key) is returned to the user that is used for login to the system. The Flowchart F1 is a pictorial representation of Algorithm A1.

The Algorithm A2 (User Registration) describes the process of user registration. User is asked for userID as input to start the registration. If userID is not found in the blockchain then the user is asked to enter name, email phone, password for registration. All user details are encrypted before storing in the blockchain that provides security for user details. Once details are stored in the blockchain successfully, the system returns a userID (public key). The userID and password (set by user) is used to login to the system. Flowchart F2 above is a graphical representation of Algorithm A2.

Algorithm A3 (User Login) illustrates the procedure for login and password recovery. Registered users can login to the system by using userID (public key) and password. If login is successful, then user_session is returned where the user is able to perform the transaction until the session expires. If the user's login is unsuccessful, then the user has to go for a password recovery process where the user is asked a series of security questions. Once the user provides correct answers, then user_session is returned. If password could not be recovered, then the user is a registered user. Flowchart F3 is a pictorial representation of the Algorithm A3.

Algorithm A4 (User Transaction) illustrates the process for performing transactions once the user has logged in successfully. Users can update product details like quality of sugarcane, location of farm, quantity, sugar mill information, payment, and other transactional data. All the information is saved in the blockchain. Once the data is stored in the blockchain, the system returns a transactionID which is used to track the product in the supply chain. The Flowchart F4 is a graphical representation of Algorithm A4.

| Algorithm-1 (A1): SugarChain | Flowchart-1 (F1): SugarChain |
|---|---|

```
START
    If new_user then
        i. Enter details (name, id, email, phone)
        ii. Return userID (public key)
    End If
    Else
        i. Enter login credentials
        If login successful then
            ii. Return user_session
        End If
            If valid user_session then
                iii. Update info
                iv. Return transaction_ID
            End If
            Else
                iii. Session expired
            End Else
        Else
            Print "Login Unsuccessful"
            ii. Go to A3
            If password recovered then
                Return user_session
            Else
                Go to A2
            End Else
        End Else
    End Else
END
```

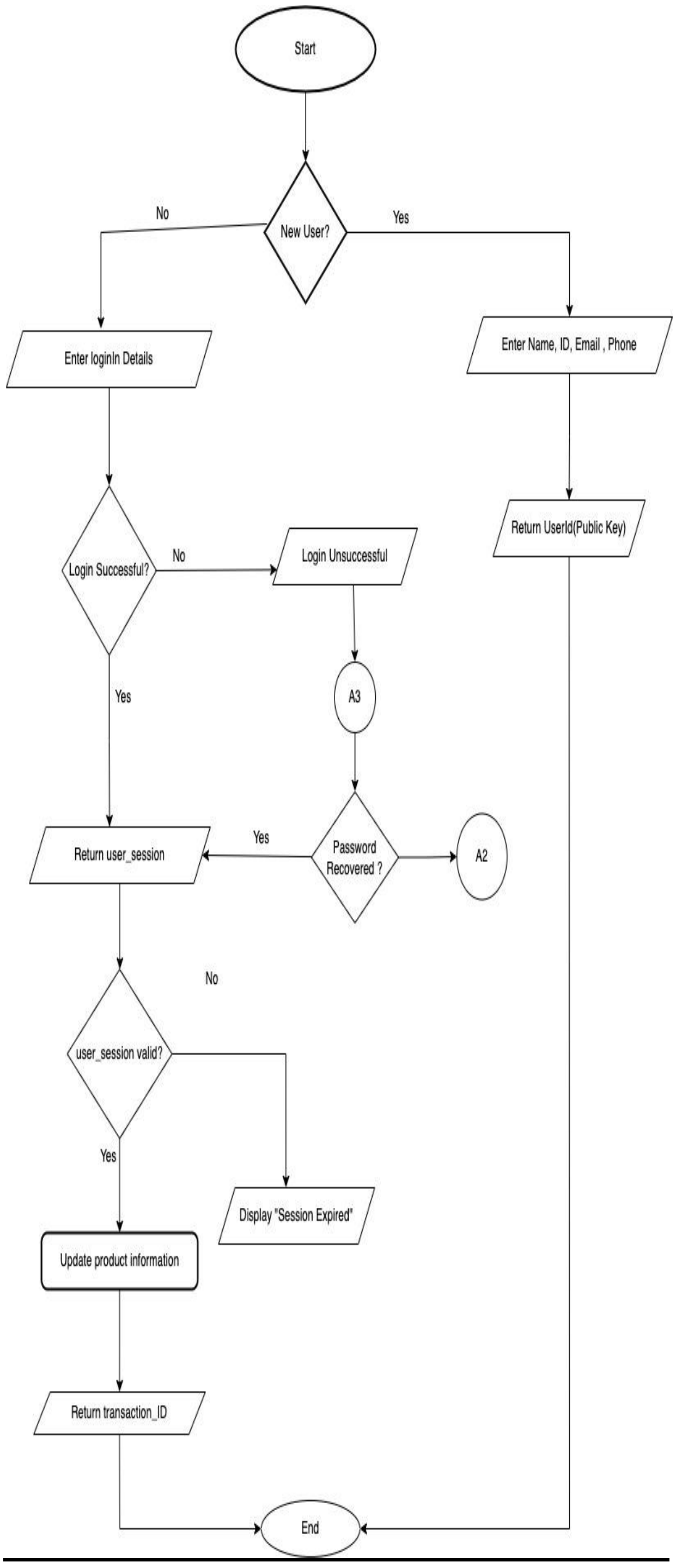

## Algorithm-2 (A2): User Registration

1. START
2. Enter userID
3. If userID not found then
4. i. Enter name, email, phone, pwd
5. ii. Encrypt all details
6. iii. Save all details in blockchain
7. iv. Return userID (public key)
8. End If
9. Else
10. i. User already exists
11. End Else
12. END

## Flowchart-2 (F2): User Registration

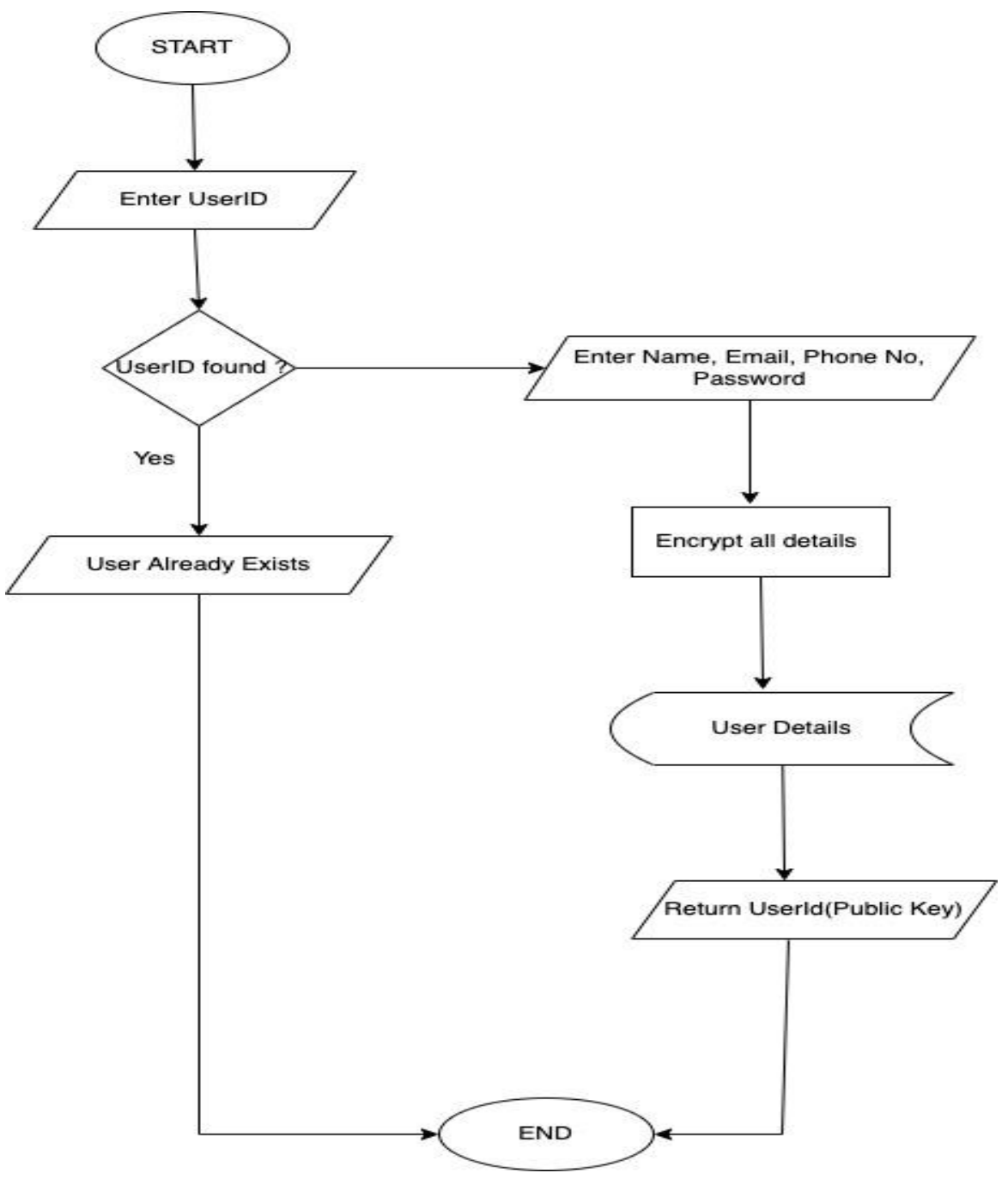

## Algorithm-3 (A3): User Login

1. START
2. Enter login credentials (userID, password)
3. If login not success
4. i. Enter secret Q's answer for pwd recovery
5. End If
6. If password recovered then
7. ii. Return user_session
8. End If
9. Else
10. iii. Unauthorized user (Not registered)
11. End else
12 Else
13. Return user_session
14. End else
15. User not registered
16. END

## Flowchart-3 (F3): User Login

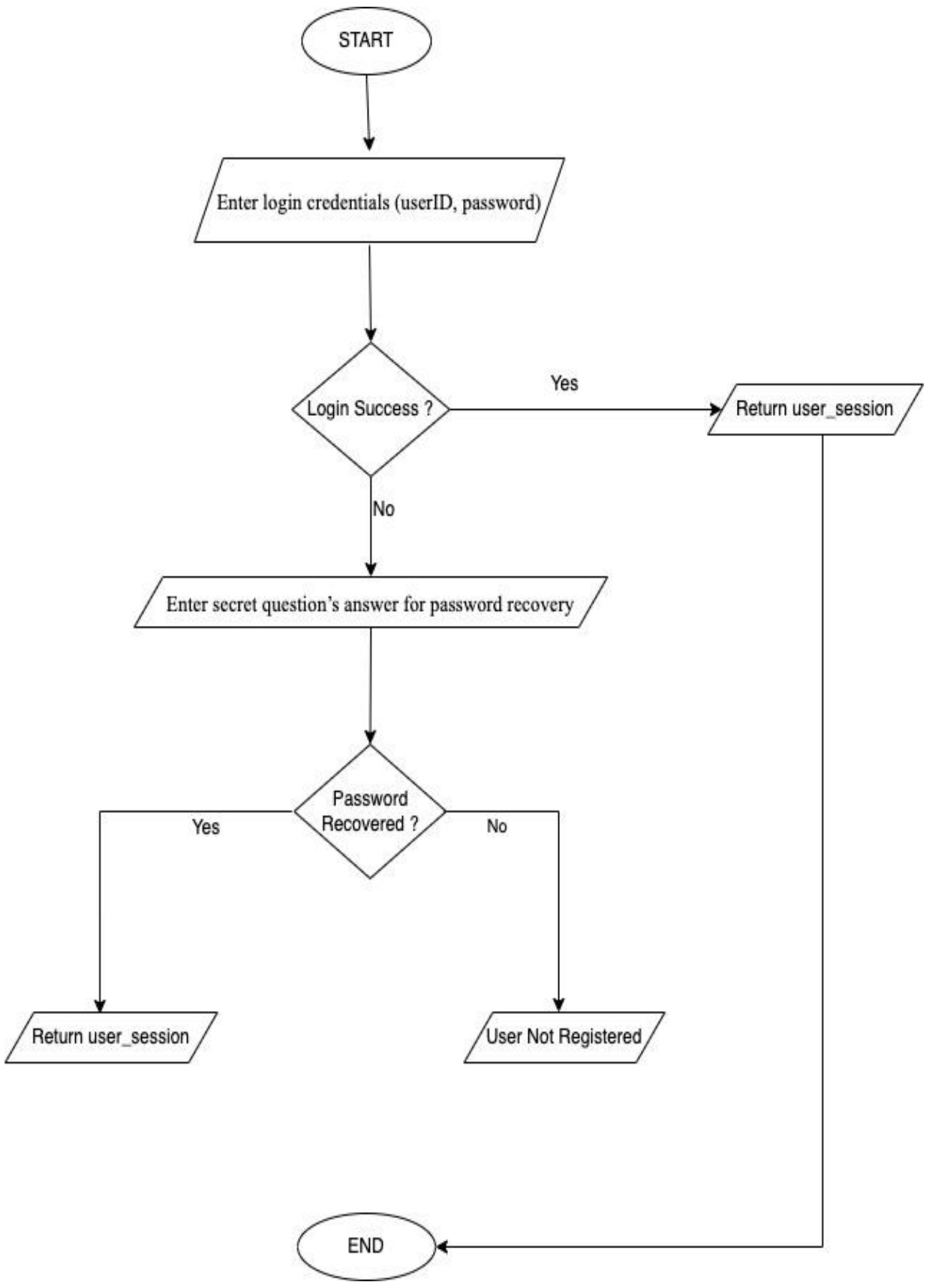

<table>
<tr><th>Algorithm-4 (A4): User Transaction</th><th>Flowchart-4 (F4): User Transaction</th></tr>
<tr><td>1. START<br>2. If valid user_session then<br>3. i. Update product (sugarcane) information (quality, quantity, location, sugar mill etc.)<br>4. ii. Save all details in blockchain<br>5. iii. Return transactionID<br>6. End If<br>7. Else<br>8. i. Session expired<br>9. End Else<br>10. END</td><td>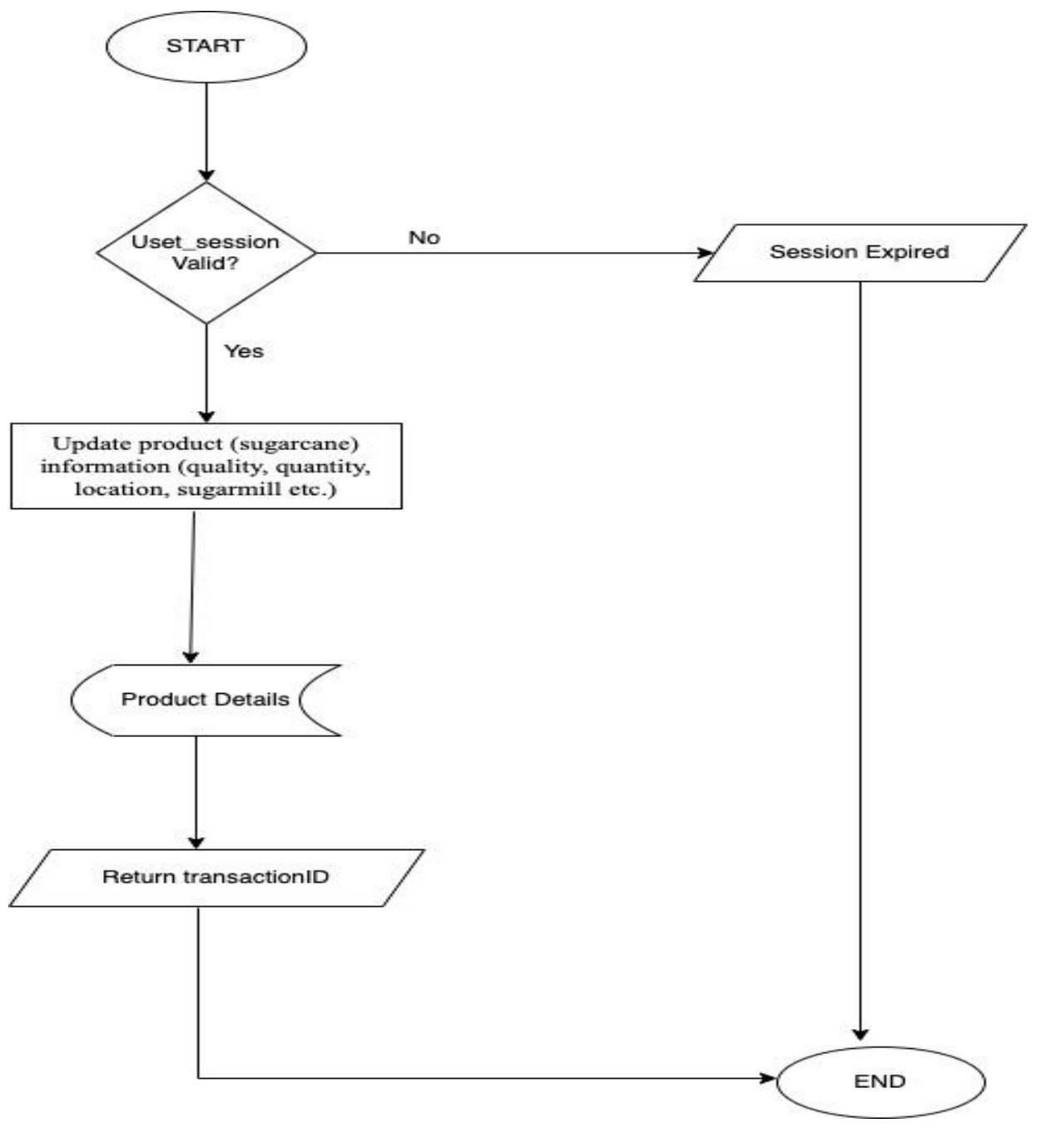
</td></tr>
</table>

## VI. PROCESS FLOW AND PROPOSED SYSTEM MODEL

From the case study presented in Section IV and the Algorithm / Flowchart of Section V, we have presented the proposed system model for our SugarChain. The BCT process flow in the figure below determines the sugarcane supply chain for our SugarChain starting from seed supplier of sugarcane to the consumer at the end. The use of distributed databases at each block is represented by the “blockchain” at the end row of the process flow. The sugarcane farmer acts as the middlemen between seed supplier and the sugar mill / factory before handing it to the distributor who purchases from sugar mills and updates his/her stocks.

The process flow of the SugarChain BCT (as shown in Figure-1 below) clearly indicates “sugarcane farmer” as the main actor, who bridges the factory with the farm. But on the contrary, a sugarcane farmer is the only one who is deprived of his actual pay and effort when we compare with “sugar mills”, distributors and retailers. The consumer is totally unaware of the process flow because the consumer has to deal only with the retailer at the end of process flow. The sugar mills or factories sell their product to distributors who in turn sells to retailers and at the end retailers sell to the consumer with appropriate profit margins.

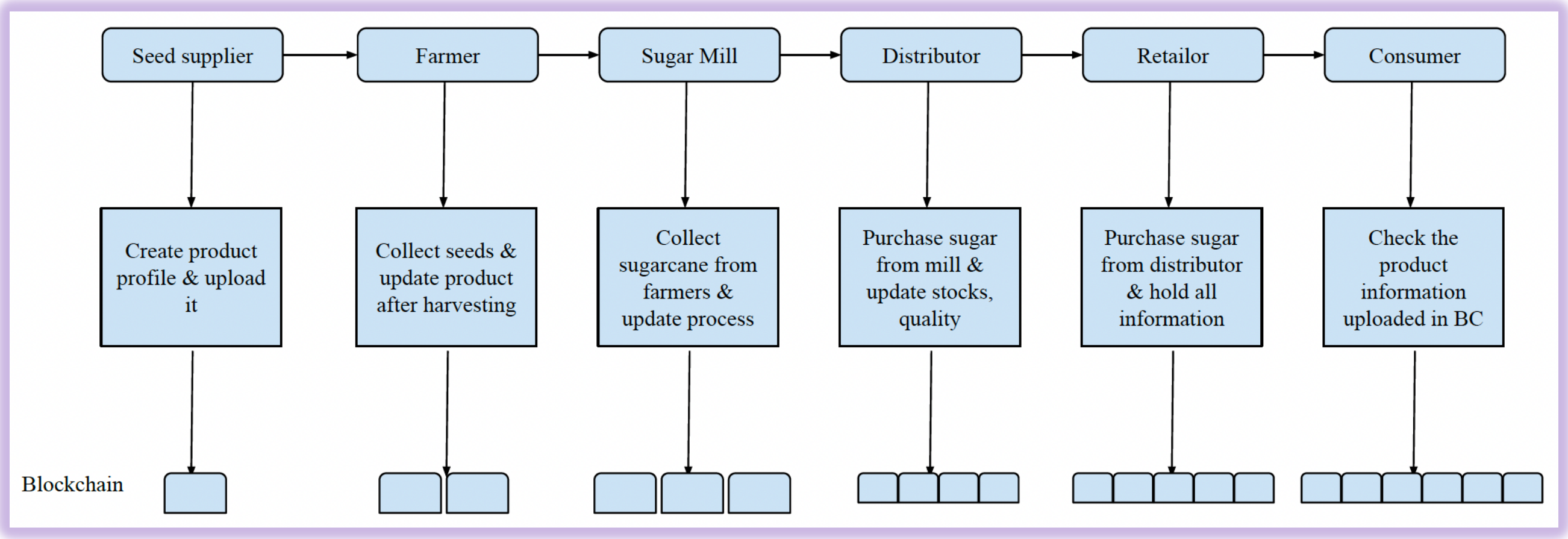

Figure-1: BCT process flow in our SugarChain *(sugarcane supply chain)* starting from seed supplier to consumer

The proposed SugarChain model is shown in Figure-2 below, which depicts sugarcane farming in conjunction with the blockchain technology. In the model, a user is a farmer and also the bearer of the wallet or the application. The application or the wallet comprises the login credentials of the farmer. The credentials is also recorded in the SugarChain network in the form of cryptographically (encrypted form) identifiers. Farmers could use the nation's backbone network to access the public services related to sugarcane farming. The important records of the sugarcane services are immutably stored in the SugarChain network meaning that the services cannot be altered or changed.

With use of the SugarChain network (as the central node of the model), we can ensure the flow of sugarcane from the farms to the sugar mills. The ensuring of the sugarcane flow will ultimately help farmers to get rid of the third-party and boost in clearance of the payments that farmers have to wait for a long period of time. The efficiency along with transparency (in purchasing and selling price of sugarcane per kg) also is improved because of the immutable recording of the sugarcane services in the SugarChain network. Although there are many other problems of farmers, ranging from irrigation to wild animals, the problem of "late payments" and "expensive transportation" at least can be overcome, with the use of the proposed system model - SugarChain.

## VII. ANALYSIS AND LIMITATION

The application of BCT in the food supply chain and agriculture has been researched as a single platform solution. We have considered it as a public platform (beside private and consortium platforms) because of farmer's agriculture / sugarcane data. The businesses that are hierarchy organized have been dominating our economy for several years. We all know that farmers are not familiar with recent technology and trends, awareness programs at the root level are needed to benefit agriculture sectors and farmers.

Our study and analysis is only focused on the problems of sugarcane farmers and the sugarcane farming. We have tried and motivated our best to collect primary data to represent the real problems and field scenarios from 40 sugarcane farmers as outlined in the Questionnaire (provided in Appendix: A1). Blockchain technology, a potentially disruptive technology, has a wise future in the food industry, supply chain management and agribusiness industry. The technology is still developing and needs more research, focused effort as well as in-depth exploitation.

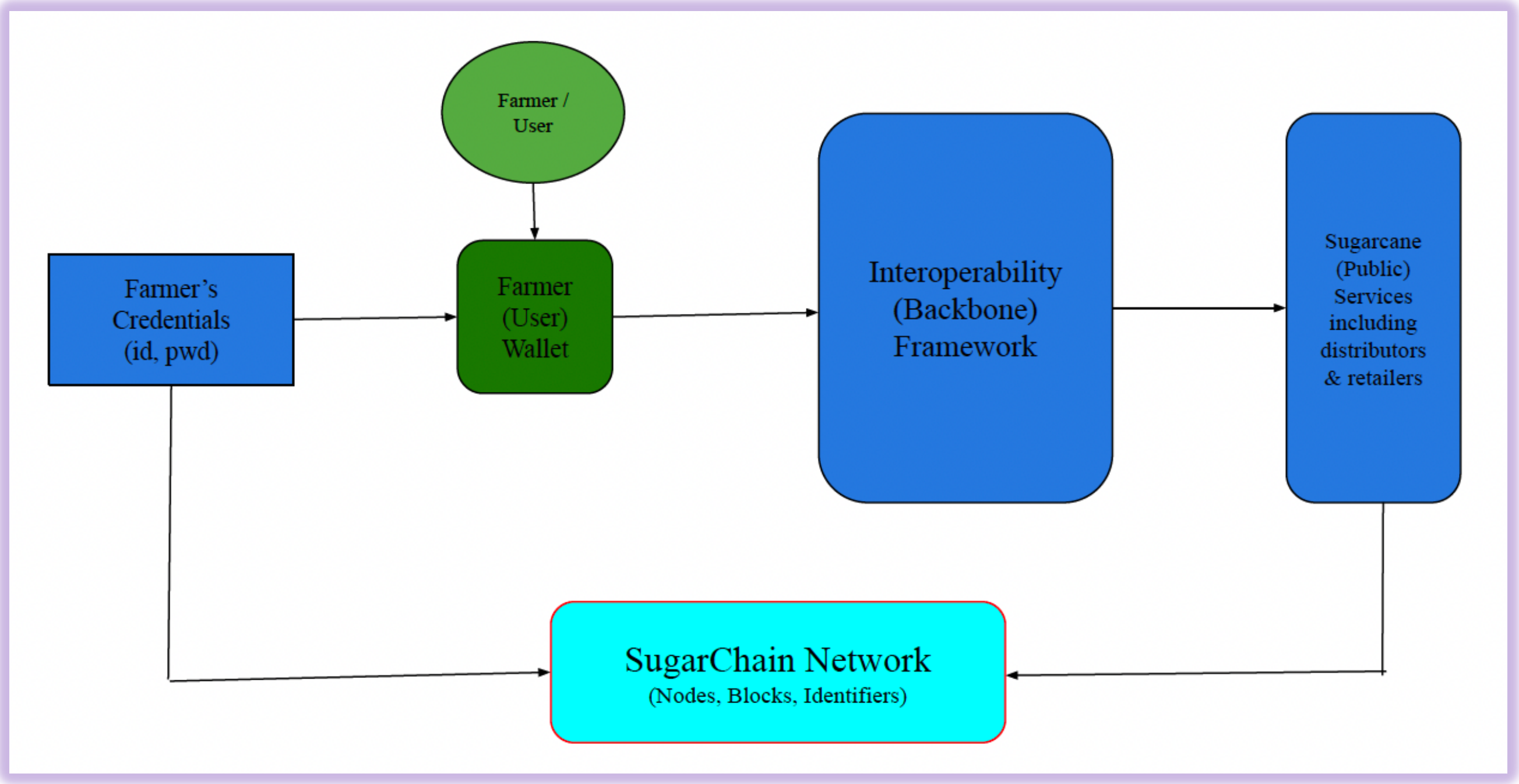

Figure-2: Proposed system model for SugarChain (sugarcane meets BCT)

## VIII. CONCLUSION

The blockchain based SugarChain scheme here tries to benefit and compensate farmers regarding their agricultural products specially sugarcane (but not limited to). The case study along with data collected from 40 farmers presented above clearly indicates that farmers are always destitute of actual pay and their labors. More than 90% of the farmers only get paid after 30 days or more of selling their sugarcane. Only 7% of farmers rely on the government's agriculture office in order to cultivate and harvest the sugarcane. Our proposed method ensures all data and details in the blockchain, where authorized users can login and access it. The user (farmer) transaction is only valid after successful return of transaction ID by registered user. We have automated the process with the sugarcane farmer’s trust and eliminate the role of middleman who is controlling the price of sugarcane.

## IX. FUTURE SCOPE

From the above presented case study, survey and proposed SugarChain model, we can clearly identify the real and burning problems of the sugarcane farmers. In the future, we want to define, frame and model several individual agriculture cases like wheat, rice, grains and other crops with the use of blockchain. Because of the double role of BCT as a ledger, public and unalterable where new farmers can be registered, and at the same time the task of managing, controlling and monitoring the proper SugarChain transaction to safeguard farmers and prevent middleman influence and scams. Blockchain is not only a database but also a computing means which can benefit farmers. It will be the worldwide interest to see how BCT can be combined with other evolving technologies (say, IoT, robotics, image fusion, etc.). The upcoming future will address how these challenges could be addressed by governments and private sectors, in order to begin blockchain technology as a secure, reliable and transparent way for the food supply chain, agricultural business and farmers.

**Credit authorship contribution statement**

**NK**: Conceptualization, Data collection, Investigation, Methodology, Formal analysis, Writing - original draft, review, and editing
**CSB**: Data collection, Investigation, Methodology, Writing - review and editing
**DK:** Data collection, Validation, Writing - review and editing
**DC:** Data collection, Validation, Writing - review and editing

**Acknowledgement**

We are thankful to all helping brains in our case study and analysis, especially those sugarcane farmers who shared their problems with us.

**Declaration of competing interest**

Authors declare no known personal relationships/competing financial interests that appeared to influence the work reported in this paper.

## Appendix

A1: Questionnaire for Sugarcane Farmers:

[Q1] For **how many years have** you been doing sugarcane farming? (a) 01- 05 years (b) 06 – 10 years (c) > 10 years
[Q2] Is sugarcane farming the **primary (major) farm** for you? (a) yes (b) no, ________ is major farm
[Q3] How **long (in months)** does it take approximately to get the sugarcane harvest? __________ months
[Q4] **Do your production cost is recovered** by selling on a price set by the government? (a) yes (b) no (c) don’t know
[Q5] At **what rate (price per kg)** do the sugar mills generally purchase your sugarcane? __________ per kg
[Q6] Is **sugarcane affected by worms/viruses**? (a) yes (b) no (c) don’t know
[Q7] Did you get paid for your sugarcane sold **instantly or after a few times**? (a) instantly (b) no after few times, ________ days
[Q8] Can you say **one major problem, while cultivating** and/or **while harvesting** the sugarcane? (a) yes _____________ (b) no
[Q9] Do you **need to find a buyer, or can you easily sell** the sugarcane? (a) find buyer (b) no need to find the buyer (c) don’t know
[Q10] Is **weather impacting/affecting sugarcane** farming? (a) yes (b) no (c) don’t know
[Q11] Is **fertilizer and seeds for sugarcane** available easily? (a) yes (b) no (c) don’t know
[Q12] How **sugarcane reaches the factory** for further processing? (a) by farmers (b) by government agencies (c) others
[Q13] Is there any **impact of wild animals** in sugarcane farming? (a) yes (b) no (c) don’t know
[Q14] Are there any **problems faced by you** (like transportations, gov. taxes or any other)? (a) yes ______ (b) no (c) don’t know
[Q15] **How do you get seeds** for cultivating / harvesting sugarcane? (a) government agriculture office (b) private seeds shop (c) others

A2: Primary data collected from farmers (based on our approved questionnaire):

1. [NK] Data-01-Farmer (24 June 2021, Thu)
2. [NK] Data-02-Farmer (25 June 2021, Fri)
3. [NK] Data-03-Farmer (26 Jun 2021, Sat)
4. [NK] Data-04-Farmer (30 Jun 2021, Wed)
5. [NK] Data-05-Farmer (02 July 2021, Fri)
6. [NK] Data-06-Farmer (03 July 2021, Sat)
7. [NK] Data-07-Farmer (06 July 2021, Tue)
8. [NK] Data-08-Farmer (06 July 2021, Tue)

15. [DK] Data-07-Farmer (28 Oct 2021, Thu)
16. [DC] Data-01-Farmer (19 Aug 2021, Thu)
17. [DC] Data-02-Farmer (19 Aug 2021, Thu)
18. [DC] Data-03-Farmer (19 Aug 2021, Thu)
19. [DC] Data-04-Farmer (19 Aug 2021, Thu)
20. [DC] Data-05-Farmer (19 Aug 2021, Thu)
21. [DC] Data-06-Farmer (30 Aug 2021, Mon)
22. [DC] Data-07-Farmer (30 Aug 2021, Mon)

29. [DK] Data-05-Farmer (22 Sep 2021, Wed)
30. [DK] Data-08-Farmer (28 Oct 2021, Thu)
31. [CSB] Data-01-Farmer (14 Aug 2021, Sat)
32. [CSB] Data-02-Farmer (14 Aug 2021, Sat)
33. [CSB] Data-03-Farmer (14 Aug 2021, Sat)
34. [CSB] Data-04-Farmer (14 Aug 2021, Sat)
35. [CSB] Data-05-Farmer (26 Aug 2021, Thu)
36. [CSB] Data-06-Farmer (29 Aug 2021, Sun)

9. [NK] Data-09-Farmer (28 July 2021, Wed)
10. [NK] Data-10-Farmer (11 Aug 2021, Wed)
11. [NK] Data-11-Farmer (20 Aug 2021, Fri)
12. [NK] Data-12-Farmer (09 Sep 2021, Thu)
13. [NK] Data-13-Farmer (21 Sep 2021, Tue)
14. [DK] Data-06-Farmer (28 Oct 2021, Thu)
23. [DC] Data-08-Farmer (30 Aug 2021, Mon)
24. [DC] Data-09-Farmer (1 Sept 2021, Wed)
25. [DK] Data-01-Farmer (12 July 2021, Mon)
26. [DK] Data-02-Farmer (16 Aug 2021, Mon)
27. [DK] Data-03-Farmer (16 Aug 2021, Mon)
28. [DK] Data-04-Farmer (17 Dec 2021, Fri)
37. [CSB] Data-07-Farmer (12 Sep 2021, Sun)
38. [CSB] Data-08-Farmer (12 Sep 2021, Sun)
39. [NK] Data-14-Farmer (08 Dec 2021, Wed)
40. [NK] Data-15-Farmer (28 Dec 2021, Tue)